\documentclass[aps, pra, reprint, superscriptaddress]{revtex4-2}

\usepackage{amsmath}
\usepackage{amssymb}
\usepackage{wasysym}
\usepackage{graphicx}
\usepackage{hyperref}
\usepackage{color}
\usepackage{siunitx}
\usepackage{xcolor}
\usepackage{changes}
\usepackage{comment}
\usepackage{bm}
\usepackage{booktabs}
\usepackage{makecell}
\usepackage{xr}
\usepackage{booktabs}
\usepackage{diagbox}
\usepackage{multirow}
\usepackage{orcidlink}

\usepackage{ulem}
\hypersetup{colorlinks=True,linkcolor=blue,citecolor=blue, breaklinks}

\allowdisplaybreaks

\begin{document}

\title{High-flux cold lithium-6 and rubidium-87 atoms from compact two-dimensional magneto-optical traps}

\author{Yun-Xuan~Lu\,\orcidlink{0009-0006-9871-3958}}
\thanks{These two authors contributed equally.}
\author{An-Wei~Zhu\,\orcidlink{0009-0001-3115-0322}}
\thanks{These two authors contributed equally.}
\author{Christine~E.~Frank\,\orcidlink{0000-0002-7372-016X}}
\author{Xin-Yi~Huang\,\orcidlink{0000-0001-8389-0093}}
\affiliation{Max-Planck-Institut f\"{u}r Quantenoptik, 85748 Garching, Germany}
\affiliation{Fakult\"{a}t f\"{u}r Physik, Ludwig-Maximilians-Universit\"{a}t, 80799 M\"{u}nchen, Germany}
\author{Xin-Yu~Luo\,\orcidlink{0000-0001-5104-3797}} 
\email[]{E-Mail: xinyu.luo@mpq.mpg.de}

\affiliation{Max-Planck-Institut f\"{u}r Quantenoptik, 85748 Garching, Germany}
\affiliation{Munich Center for Quantum Science and Technology, 80799 M\"{u}nchen, Germany}

\date{\today}

\begin{abstract}
We report a compact setup with in-series two-dimensional magneto-optical traps (2D MOTs) that provides high-flux cold lithium and rubidium atoms. Thanks to the efficient short-distance Zeeman slowing, the maximum 3D MOT loading rate of lithium atoms reaches a record value of $6.6\times 10^{9}$ atoms/s at a moderate lithium-oven temperature of 372 degrees Celsius, which is 44 times higher than that without the Zeeman slowing light. The flux of rubidium is also as high as $2.3\times10^9$ atoms/s with the rubidium oven held at room temperature. Meanwhile, the entire vacuum-chamber system, including an ultra-high-vacuum science cell, is within a small volume of $55\times65\times70~\mathrm{cm}^3$. Our work represents a substantial improvement over traditional bulky and complex dual-species cold-atom setups. It provides a good starting point for the fast production of a double-degenerate lithium-rubidium atomic mixture and large samples of ultracold lithium-rubidium ground-state molecules.
\end{abstract}

\maketitle
\section{Introduction}\label{Introduction}

Ultracold polar molecules have emerged as a powerful platform for studying ultracold chemistry, testing fundamental symmetry, and exploring strongly correlated many-body systems with strong dipole-dipole interactions~\cite{Carr2009,Ye2017}. Among bialkali polar molecules, $^6$Li$^{87}$Rb ground-state molecules stand out as they exhibit the second-largest electric dipole moment~\cite{yin2024stirap}. This makes it a promising candidate for realizing and probing novel quantum phases, such as the long-sought dipolar Bardeen-Cooper-Schrieffer to Bose-Einstein condensate (BCS-BEC) crossover~\cite{deng2023}. So far, several labs have successfully produced degenerate mixtures of fermionic lithium and bosonic rubidium atoms~\cite{silber2005,taglieber2008} and observed Feshbach resonances between lithium and rubidium atoms~\cite{deh2008,deh2010}.  Ultracold $^6$Li$^{87}$Rb polar molecules remain elusive in experiments.

Next-generation experiments with ultracold dual-species atoms and polar molecules require shorter cycle times and miniaturized system volume, thereby demanding a compact dual-species cold-atom source with high atomic flux.
In the early days, achieving high flux often necessitated complex setups with multiple bulky vacuum chambers, especially for high-melting-point elements such as lithium and sodium~\cite{phillips1982Na, joffe1993Na, mewes1999_Li,hadzibabic2002, taglieber2006, marti2010LiRb, ridinger2011LiK, park2012, hopkins2016, wu2017LiK, ilzhofer2018, ma2019, Hulet2020_Li_QGM, warner2021, vayninger2025}. Those elements typically require long Zeeman slowers to decelerate fast atomic beams from hot ovens. While compact, two-dimensional magneto-optical traps (2D MOTs) have relatively low capture velocities, therefore tend to work more efficiently for low-melting-point elements such as rubidium and cesium~\cite{dieckmann1998a,lam2020, chenLithiumcesiumSlowBeam2021, pur2023,sutradhar2023,matthies2024}. In 2009, Tiecke \textit{et al}. demonstrated a compact lithium 2D MOT using permanent magnets, achieving loading rates comparable to those of Zeeman slowers~\cite{tiecke2009}. Recently, a modularized lithium quantum gas setup based on a 2D MOT has been developed in Heidelberg and Munich~\cite{hammel2025, jain2025_unirand_li}, which further enhances the loading rate by employing high-power and top-hat cooling beams~\cite{hammel2025}.
Combining a 2D MOT with a short-distance Zeeman slower can substantially enhance its capture velocity and thereby loading rates. This has been demonstrated for sodium, strontium, and ytterbium atoms~\cite{lamporesi2013, li2020, li2023, nosske2017, wodey2021}, but not for lithium atoms yet. Moreover, integrating high- and low-melting-point elements into a single compact system while maintaining high flux for both species remains an experimental challenge.

In this work, we present a compact and high-flux dual-species cold-atom source that overcomes these challenges by combining a rubidium 2D MOT in series with a lithium 2D MOT, enhanced by a short-distance Zeeman slower for lithium. As a result, the 3D MOT loading rate of lithium atoms reaches a record value that is 44 times higher than that without the Zeeman slowing light. The flux of rubidium also reaches a reasonably high value. Meanwhile, the vacuum chamber is so small that it can be mounted on a movable breadboard~\cite{hammel2025}. Thereby, we can smoothly extract it from the magnetic coils and the forest of optics for laser beam alignment or maintenance, then push it back to the original position. Thanks to the in-series 2D MOTs configuration, the compact vacuum system also allows more than 270 degrees wide optical access to the science cell in the horizontal plane, providing an excellent starting point for future experiments.

The paper is organized as follows: in Sec .~\ref{ExperimentalSetup}, we describe the experimental setup, including the vacuum system, laser sources, and MOT configurations. In Sec .~\ref{sec:results_and_discussions}, we present the loading behavior of the dual-species MOT and detail the optimization procedures. In Sec .~\ref{sec:summary}, we summarize the main results and discuss the implications of our work for future experiments.

\begin{figure*}[htbp]
    \centering
    \includegraphics{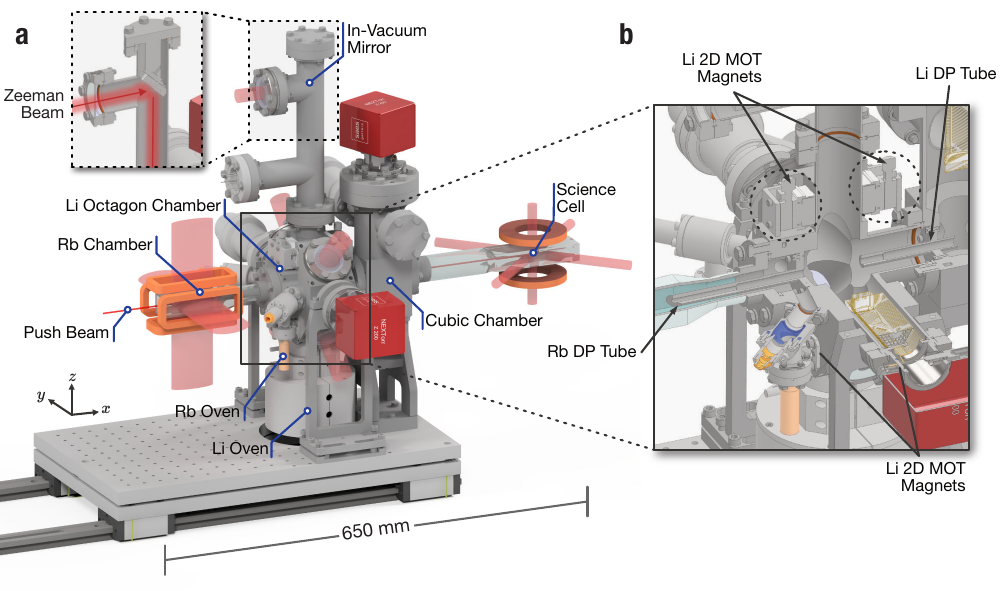}
    \caption{Lithium-Rubidium Dual-Species Vacuum Setup. (a): Overview of the vacuum system with laser beams for the Rb 2D MOT (left), the Li 2D MOT (middle), and the dual-species 3D MOT (right) and the short-distance Li Zeeman slower (up). The miniaturized vacuum system has a dimension of $55\times65\times70\ \mathrm{cm}^3$. The vacuum setup sits on a rail system, which allows for 1D translation along the $x$ direction for easy vacuum diagnostics and optics alignment. A gold-coated aluminum mirror (\#47-116, Edmund optics) is placed $300\ \mathrm{mm}$ above the center of the octagon chamber to limit the consequences of hot atom flux directly shooting towards the cold viewport and coat the viewport to compromise the transmission. (b): Three-quarter view of the octagon chamber section, where two stages of differential pumping (DP) channels are shown. Four stacks of magnets are attached to the side of the Li octagon chamber in a way that generates a quadruple magnetic field. The center of the magnetic field can be fine-tuned using the translation stages to align the 2D MOT with the center of the DP tube.}
    \label{fig:figure1}
\end{figure*}
\section{Experimental setup}\label{ExperimentalSetup}
\subsection{Vacuum system and general setup}\label{VacuumSystem}

Our vacuum system integrates Rb and Li two-dimensional magneto-optical traps (2D MOTs) arranged in series inside a high-vacuum (HV) region. These sources generate high-flux pre-cooled atomic beams that are further collimated by push beams and captured in a dual-species 3D MOT located in an ultra-high-vacuum (UHV) science cell, as shown in Fig.~\ref{fig:figure1}. Two differential pumping channels (DP) maintain a pressure drop of more than three orders of magnitude between the HV and UHV regions while preserving large solid angles for both beams (see Appendix~\ref{sec:appendix_DP}). Placing both 2D MOTs on the same side of the main chamber maximizes optical and microwave access to the science cell for manipulation of ultracold atoms and molecules~\cite{biswas2025_nak_microwave}. During daily operation, we measure a pressure on the order of $1\times10^{-9}\ \mathrm{mbar}$ in the Li octagon chamber and  $3\times10^{-11}\ \mathrm{mbar}$ in the cubic science chamber. This pressure is consistent with the measured vacuum-limited lifetime exceeding $50\ \mathrm{s}$ in the science chamber.

The Rb 2D MOT loads atoms directly from the background vapor at a pressure of about $10^{-8}\ \mathrm{mbar}$ and provides an atomic flux of roughly $10^{10}\ \mathrm{atoms/s}$ in a $2\mathrm{D}^+$ configuration. In contrast, $^6\mathrm{Li}$ has an extremely low vapor pressure at room temperature ($\sim 10^{-20}\ \mathrm{mbar}$)~\cite{gehm2003} and therefore requires an oven-based source. We place the high-vapor-pressure, low-divergence Rb 2D MOT at the upper stream of the vacuum system, while the low-vapor-pressure, high-divergence Li 2D MOT is located in the middle, with the crucial UHV science chamber further downstream. The Li 2D MOT chamber also serves as an additional differential pumping stage between the Rb 2D MOT region and the science chamber. This layout facilitates a natural HV-to-UHV transition, allowing for the independent optimization of the spatially separated Rb and Li 2D MOTs, thereby enabling the simultaneous high-flux loading of both species into the 3D MOT.

To achieve sufficient Li vapor pressure for trapping, the oven is heated to about $350\ \mathrm{^\circ C}$, which increases atomic density but also shifts the velocity distribution toward higher speeds. We therefore employ a 2D MOT~\cite{tiecke2009} enhanced by a short-distance Zeeman slower. The Li 2D MOT and Zeeman slower share the magnetic field of four stacks of permanent magnets mounted on the octagon chamber (Fig.~\ref{fig:figure1}b), following approaches demonstrated for Na~\cite{lamporesi2013} and Sr~\cite{nosske2017}. A Zeeman beam is reflected by an in-vacuum mirror (see Fig.~\ref{fig:figure1}a) to counter-propagate against the atomic flux and decelerate atoms emerging from the oven.

UHV conditions are maintained using two ion-getter pumps (\texttt{NEXTorr\textsuperscript{\textregistered} Z200}, SAES). One pump is attached to the octagon chamber, evacuating the Li chamber, the double-T section, and the Rb chamber through a low-conductance DP channel that limits the influence of Rb vapor. A second pump on the cubic chamber maintains the UHV in the science cell.

\subsection{Source and Ovens}\label{Ovens}
We filled the Li oven with 10 gram chunks of enriched $^{6}\mathrm{Li}$ metal source ($95\%$ abundance, CAS:14258-72-1, \#340421-10G, Merck KGaA, Darmstadt), that were pre-cleaned with cyclohexane ($\geq99.5\%$ (GC), Sigma-Aldrich) in order to remove the mineral oil. The cleaning and oven filling were carried out in the air and completed in about 1.5 hours. The Li oven was then baked on a separate vacuum setup at $470\ \mathrm{^\circ C}$ for one hour. This procedure eliminated residual oil contaminants and hydrogen that could compromise the vacuum quality. The Li oven was subsequently connected to the octagon chamber via a custom CF 25 flange secured with six screws to ensure superior vacuum sealing. The oven features a deliberately thin neck design to minimize thermal conduction to the octagon chamber. To prevent clogging of the oven tube, a hollow tungsten tube with high thermal conductivity extends from the bottom of the oven up to the oven opening~\footnote{Private communication with C. Gross}. The geometric configuration places the oven bottom $190\ \mathrm{mm}$ from the center of the octagon chamber and $495\ \mathrm{mm}$ from the in-vacuum mirror.

The Rb source (CAS: 7440-17-7, CAT: 044214.03,  Thermo Fisher Scientific Inc.) consists of 1 gram of natural isotopic abundance rubidium ($\eta_{^{87}\mathrm{Rb}} = 27.83(2)\%$ and $\eta_{^{85}\mathrm{Rb}} = 72.17(2)\%$~\cite{daniela.steck2023, daniela.steck2023a}), sealed in a glass ampoule to prevent oxidation. We designed the Rb oven using a copper tube with a specialized opening that serves as a gasket (LIMIT VACUUM, Beijing). This assembly is sealed with a custom flange and connected to the Rb chamber via an angle valve, allowing for a controlled and gentle introduction of the atomic vapor. To activate the Rb source, we mechanically crush the copper tube, breaking the enclosed ampoule during the final stage of the vacuum baking process.

\subsection{Dual species laser system}
\begin{figure}[t]
\centering
\includegraphics{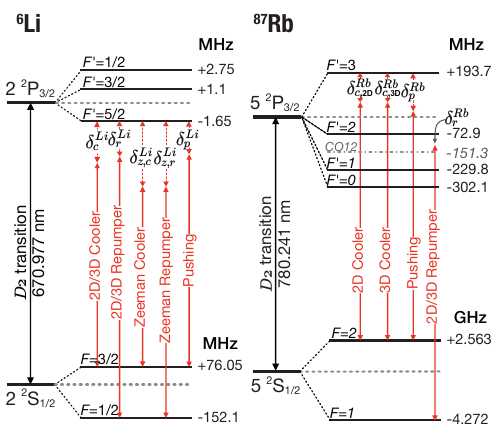}
\caption{\label{fig:figure2}
Laser frequency scheme (not to scale).
Left: $^6\mathrm{Li}$ $D_2$ transition from the $2^2\mathrm{S}_{1/2}$ state to the $2^2\mathrm{P}_{3/2}$ manifold.
Red arrows indicate the corresponding laser frequencies, and $\delta^\mathrm{Li}_i$ denote the detunings of the individual beams with respect to the upper manifold.
Transition frequencies are taken from~\cite{gehm2003}.
Right: $^{87}\mathrm{Rb}$ $D_2$ transition from the $5^2\mathrm{S}_{1/2}$ state to the $5^2\mathrm{P}_{3/2}$ state.
Red arrows indicate the corresponding laser frequency, and $\delta^\mathrm{Rb}_i$ denote the detunings of the individual beams with respect to the corresponding transition.
Transition frequencies are taken from~\cite{daniela.steck2023}.
}
\end{figure}

Leveraging the strong and well-characterized $\mathrm{S} \to \mathrm{P}$ transitions of alkali metal elements, we implement optical trapping and cooling in our MOT using the $D_2$ transitions of both $^{6}\mathrm{Li}$ (671 nm) and $^{87}\text{Rb}$ (780 nm). The laser frequencies are tuned relative to their hyperfine structures, as illustrated in Fig.~\ref{fig:figure2}.

\subsubsection{Lithium laser system}\label{subsection:li_laser_system}
The $^{6}\text{Li}-D_2$ transition possesses a natural linewidth of $\Gamma_{\mathrm{Li}}/2\pi \approx 5.87\ \mathrm{MHz}$, while the hyperfine splitting between the upper manifold $|2^2\text{P}_{3/2}; F'=1/2, 3/2, 5/2\rangle$ is merely $4.4\ \mathrm{MHz}$~\cite{das2007, gehm2003}. While driving the cycling transition $|2^2\text{S}_{1/2}; F=3/2\rangle \to |2^2\text{P}_{3/2}; F'=5/2\rangle$ for laser cooling, the transitions to the $|2^2\text{P}_{3/2}; F'=1/2, 3/2\rangle$ states are simultaneously driven. Consequently, no convenient fully closed cooling transitions can be utilized here~\cite{tiecke2009}. 

For efficient trapping and cooling, an intense 'repumping' beam ($|2^2\text{S}_{1/2}; F=1/2\rangle \to |2^2\text{P}_{3/2}; F'\rangle$) is required. Its intensity ought to be comparable to that of the 'cooling' beam ($|2^2\text{S}_{1/2}; F=3/2\rangle \to |2^2\text{P}_{3/2}; F'\rangle$) to suppress atom loss caused by population falling into the dark state.

For the short-distance Zeeman slower (ZS), we implement a ZS cooling beam addressing the $|2^2\text{S}_{1/2}; F=3/2\rangle \to |2^2\text{P}_{3/2}; F'\rangle$ transition and a ZS repumping beam targeting the $|2^2\text{S}_{1/2}; F=1/2\rangle \to |2^2\text{P}_{3/2}; F'\rangle$ transition, operating in coordination with the tailored magnetic field profile and induced Zeeman shift. To maximize the atomic flux directed toward the 3D MOT, we additionally employ a pushing beam that is slightly red-detuned from the $|2^2\text{S}_{1/2}; F=3/2\rangle \to |2^2\text{P}_{3/2}; F'\rangle$ transition, which effectively collimates the atomic beam along the axial direction.

A schematic of the lithium $D_2$ line laser system is presented in Fig.~\ref{fig:figure3}. The system is powered by a $2\ \mathrm{W}$ laser operating at $671\ \mathrm{nm}$ (Precilasers, Shanghai), which provides sufficient optical power for both cooling and Zeeman slowing. This laser is offset-locked via an optical phase lock loop (OPLL) to a reference laser, which itself is locked to the $|2^2\text{S}_{1/2}; F=1/2\rangle \to |2^2\text{P}_{3/2}; F'\rangle$ and $|2^2\text{S}_{1/2}; F=3/2\rangle \to |2^2\text{P}_{3/2}; F'\rangle$ crossover transition of $\mathrm{^6Li}$ via saturated absorption spectroscopy. Multiple double-pass acousto-optic modulators (AOMs) are utilized to independently tune each beam to its required detuning $\delta^\mathrm{Li}$. For the Zeeman slower branch specifically, we incorporate a tunable resonant electro-optical modulator (EOM) after the AOM to generate sidebands at approximately $228\ \mathrm{MHz}$ from the carrier frequency, which serves as the ZS repumping beam. The relative intensity between the carrier and the sidebands can be controlled by adjusting the RF power delivered to the EOM.

\subsubsection{Rubidium Laser System}
The laser frequency scheme for laser cooling of $^{87}\text{Rb}$ is shown in Fig.~\ref{fig:figure2}, following many of the well-established Rb cooling schemes, for example~\cite{wei2020}. In contrast to $^6\text{Li}$, the $^{87}\text{Rb}\ D_2$ transition exhibits a natural linewidth of $\Gamma_{\mathrm{Rb}}/2\pi \approx 6.07\ \mathrm{MHz}$, with hyperfine splittings in the $5^2\mathrm{P}_{3/2}$ states ranging from tens to hundreds of $\mathrm{MHz}$~\cite{daniela.steck2023, ye1996}. These well-resolved energy separations significantly simplify the implementation of laser cooling. For our cooling scheme, we utilize the closed $|5^2\mathrm{S}_{1/2}; F=2\rangle \to |5^2\mathrm{P}_{3/2}; F'=3\rangle$ transition as the primary cooling cycle, supplemented by a repumping beam resonant with the $|5^2\mathrm{S}_{1/2}; F=1\rangle \to |5^2\mathrm{P}_{3/2}; F'=2\rangle$ transition to recover atoms that decay out of the cooling cycle. Analogous to our lithium configuration, we implement a push beam, red-detuned from the $|5^2\mathrm{S}_{1/2}; F=2\rangle \to |5^2\mathrm{P}_{3/2}; F'=3\rangle$ transition, to enhance the directed atomic flux along the axial direction.

\begin{figure}[t]
\centering
\includegraphics{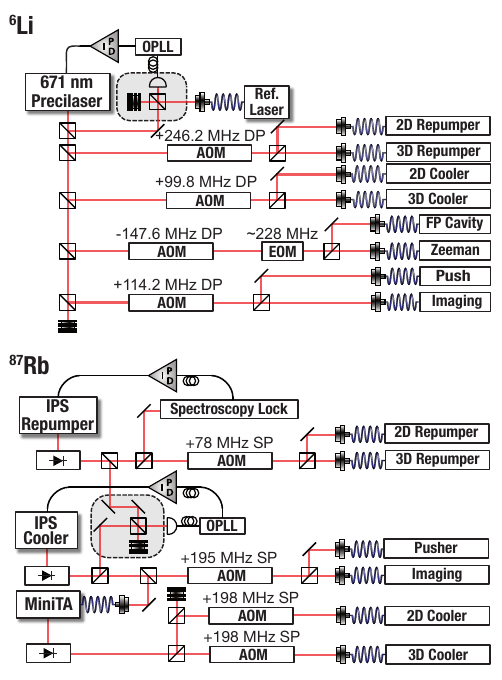}
\caption{\label{fig:figure3}
Schematic of the lithium and rubidium laser systems.
Acousto-optic modulators (AOMs) in double-pass (DP) or single-pass (SP) configurations are used to modulate the laser frequencies.
In the Li laser system, a tunable electro-optical modulator (EOM) in the Zeeman-slower beam path generates sidebands at approximately $228\,\mathrm{MHz}$ from the carrier to serve as the repumping beam.
A Fabry–Perot cavity monitors the sideband frequency and intensity.
}
\end{figure}

A detailed schematic of the Rb laser system is depicted in Fig.~\ref{fig:figure3}. Two IPS (Innovative Photonic Solutions, Plainsboro) laser diodes centered at $780.2\ \mathrm{nm}$ serve as seed lasers. One of these is locked to the $|5^2\mathrm{S}_{1/2}; F=1\rangle \to |5^2\mathrm{P}_{3/2}; F'=1\rangle$ and $|5^2\mathrm{S}_{1/2}; F=1\rangle \to |5^2\mathrm{P}_{3/2}; F'=2\rangle$ crossover peak via saturated absorption spectroscopy. This diode also provides the repumping light after modulation by single-pass AOMs.

The second laser diode is offset-locked to the repumping seed laser via an OPLL, maintaining a frequency offset of approximately $-6.8\ \mathrm{GHz}$ relative to the master laser. Both the push beam and imaging beam are derived directly from the seed laser output. A portion of this output ($\sim20\ \mathrm{mW}$) seeds a tapered amplifier (miniTA, EAGLEYARD Photonics GmbH, Berlin), which boosts the optical power to $2\ \mathrm{W}$. Multiple AOMs subsequently modulate the amplified output to establish the required detunings for each cooling beam.

\subsection{The 2D MOT configuration}

The Rb 2D MOT is housed in a glass cell chamber, ensuring excellent optical access for the large laser beams that capture as many atoms from the Rb vapor. The cooling and repumping beams are combined using a $2\times2$ polarization-maintaining (PM) fiber coupler (PN780R5A2, Thorlabs Inc.) and collimated by optics on a breadboard around the chamber. A retro-reflection configuration is employed to maximize the use of available laser power. A telescope with elliptical lenses expands the beams to achieve $1/e^2$ waists of $40 \times 9.4\ \mathrm{mm}$ to accommodate the geometry of the glass cell.

To generate the magnetic field for magneto-optical trapping, two pairs of rectangular coils are configured in an anti-Helmholtz arrangement, allowing for fine adjustment of the field center in the $yz$-plane to align the 2D MOT with the push beam and DP tube. We apply a gradient of about  $\partial B_y/\partial y = 9.5\ \mathrm{G/cm}$ for the Rb 2D MOT trapping.

The Li 2D MOT is prepared in a metal octagon chamber with two horizontal openings for the ion pump and an angle valve for baking. The atomic flux from the Li oven enters through an opening at the bottom and is pre-cooled by a Zeeman slower (ZS) beam. Four 1-inch viewports on the octagon chamber provide optical access for large trapping beams. At the center of the chamber, four laser beams (two pairs each in retro-reflection configurations) intersect to provide transverse confinement. Each beam has a $1/e^2$ waist of $\sim14\ \mathrm{mm}$.

The magnetic field for the Li 2D MOT is generated by four stacks of neodymium permanent magnets (N750-RB, Eclipse Magnetics Ltd, Sheffield) positioned at coordinates $(\pm l_x/2=35~\mathrm{mm}, 0, \pm l_z/2=53.5~\mathrm{mm})$ relative to the octagon chamber center (see Fig.~\ref{fig:figure1}b). Each stack consists of 10-12 individual magnets, each with dimensions of $10 \times 3 \times 25\ \mathrm{mm^3}$ and magnetization directed along the 3 mm edge.  The magnet stacks are housed in PEEK holders and mounted on 1D translation stages to allow precise adjustment of the field center position.

\subsection{Li short-distance Zeeman slowing design}\label{section:zs_design}
Prior to vacuum chamber fabrication, we conducted extensive numerical simulations to optimize the distances ($l_x$ and $l_z$) between the magnets stacks and the number of magnets per stack, with the goal of maintaining a magnetic field gradient of approximately $50\ \mathrm{G/cm}$ required for efficient 2D MOT operation, as detailed in Appendix~\ref{sec:appendix_ZS}. 

The simulations were conducted based on the simplified models outlined in~\cite{ tiecke2009,lamporesi2013,nosske2017}. The total atomic flux emitted through the oven aperture (area $A$) is given by:
\begin{equation}
    \Phi_\mathrm{tot} = \frac{1}{4} n \overline{v} A,
\end{equation}
where $n$ is the atomic number density and $\bar v=\sqrt{8k_B T/(\pi m)}$ is the mean thermal speed at oven temperature $T$. From this flux, only a fraction can be captured by the 2D MOT:
\begin{equation}
\Phi_c \approx \frac{1}{2} a_6 n \overline{v} A \left(\frac{v_c}{\alpha}\right)^4 \frac{\Omega_c}{4\pi}.
\end{equation}
This formula is a good approximation for the capture velocity $v_c \ll \alpha$\cite{tiecke2009}. Here, $a_6$ is the isotopic abundance of $^6\text{Li}$, $\alpha = \sqrt{2k_B T/m}$ is the most probable velocity, and $\Omega_c$ is the solid angle subtended by the 2D MOT region. The oven temperature $T$ determines the velocity distribution and the saturated vapor pressure $p_s$, which is derived from the Clausius-Clapeyron relation. Further details can be found in~\cite{tiecke2009}.

\begin{figure}[t]
\centering
\includegraphics{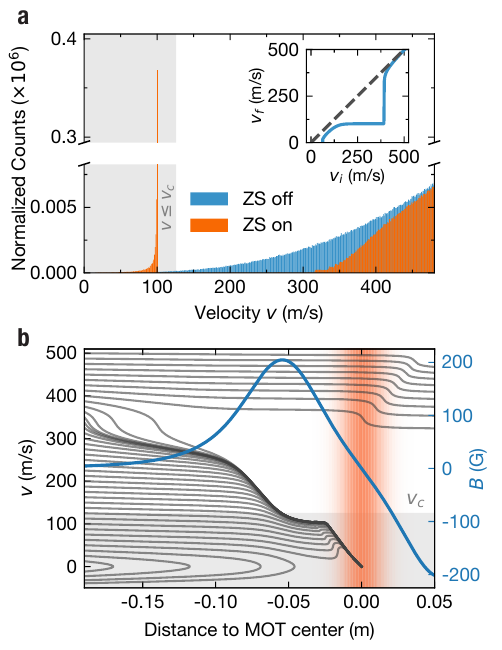}
\caption{\label{fig:figure4}
Simulation of the short-distance Li Zeeman slowing and 2D MOT.
(a) Velocity distribution of $10^6$ Li atoms with and without Zeeman slowing. The Zeeman slower reshapes the initial distribution by decelerating atoms in the blue region into the orange peak, bringing them below the 2D MOT capture velocity $v_c$.
Inset: velocity deformation caused by Zeeman slowing (figure inspired by~\cite{lamporesi2013}).
(b) Numerical simulations of atomic trajectories along the applied magnetic field.
The 2D MOT beam region is shown in orange.
The light gray area indicates the capture velocity $v_c$ of the 2D MOT.
}
\end{figure}

The Zeeman slower deforms the atomic velocity distribution, significantly increasing the fraction of atoms with velocities below the capture velocity of the MOT ($v_\mathrm{end} \leq v_{c,\mathrm{MOT}}$), as illustrated in Fig.~\ref{fig:figure4}a. Typically, a ZS system comprises two essential components: (i) a varying magnetic field along the atomic trajectory to induce position-dependent Zeeman shifts of the atomic energy levels, and (ii) a counter-propagating laser beam for controlled deceleration. In this design, we leverage the long tail of the existing magnetic field generated by the 2D MOT magnets, requiring only the addition of a dedicated Zeeman slower laser beam to complete the slowing mechanism.

To minimize contamination of the glass viewport caused by lithium's high sticking coefficient and small mass, the Zeeman beam is redirected using a gold-coated in-vacuum mirror mounted $300\ \mathrm{mm}$ above the center of the octagon chamber on the top of a so-called double-T structure (compare Fig.~\ref{fig:figure1}). However, we still observe some contamination after several months of operation, although the ZS performance is barely affected. An alternative approach would be to heat a viewport made of sapphire crystal positioned directly against the atomic flux, as demonstrated by Nosske \textit{et al.}~\cite{nosske2017}. Their implementation also showed degradation in optical transmission over a period of several months. The ZS beam is far-detuned (approximately $-85\Gamma$) to take advantage of the positive gradient region of the magnetic field ($\partial B_y/\partial z > 0$), minimizing interference with the 2D MOT operation. In contrast to conventional ZS designs~\cite{phillips1982Na}, our system employs a quadrupole magnetic field oriented transverse to the atomic beam trajectory. This configuration demands circularly polarized light relative to the local magnetic field vector. Consequently, when using a linearly polarized ZS beam (as in our setup), only the $\sigma^-$ polarization component effectively contributes to the slowing process, reducing the practical intensity of the ZS beam by 50\%~\cite{nosske2017}. Along one dimension, the radiation force acting on the atoms is described by:
\begin{equation}
\label{eqn:zeeman_slower_radiation_force}
F_{-} = -\frac{\hbar k \Gamma}{2} \frac{S_0}{1 + S_0 + \left(2\delta_{-}/\Gamma\right)^2},\quad a = \frac{F_{-}}{m},
\end{equation}
where $S_0 = I/I_\mathrm{sat}$ is the saturation parameter, $\delta_{-}$ is the effective detuning, and $\Gamma$ is the natural linewidth of the transition.  The effective detuning is given by:
\begin{equation}
\label{eqn:zeeman_slower_detuning}
\delta_{-} = \delta_0 + k v - \frac{\mu' \eta_B B}{\hbar}.
\end{equation}
Here, $\delta_0$ is the zero-field detuning, $v$ is the atomic velocity, $B$ is the magnetic field strength, and the factor $0 < \eta_B < 1$ accounts for shielding effects of the metal chamber to the magnetic field\cite{lamporesi2013}.

The motion of atoms along the $z$-axis is numerically simulated using the equation of motion derived from the radiation force. The capture velocity $v_c$ affects the gain $G$ of the  ZS according to:
\begin{equation}
    G = \frac{\displaystyle \int_0^{v^*} v_i^3 \exp\left[-v_i^2 / \alpha\right]\,\mathrm{d} v_i}{\displaystyle \int_0^{v_c}   v_i^3 \exp\left[-v_i^2 / \alpha\right]\, \mathrm{d} v_i},
\label{eqn:zeeman_slower_gain}
\end{equation}
where $v^*$ is defined as the maximum initial velocity such that the final slowed velocity satisfies $v_f(v_i) \leq v_c$. The integrals are weighted by the initial thermal flux distribution, where $\alpha$ is related to the most probable velocity. This ratio $G$ compares the total flux of atoms capturable by the trap with and without the operation of the Zeeman slower.

Our simulation suggests a gain factor of 53. These simulations, as depicted in Fig.~\ref{fig:figure4} b and detailed in the Appendix~\ref{sec:appendix_ZS}, guided the final design of the Zeeman slower to maximize the atomic flux reaching the 2D MOT region.

\subsection{Push Beam}
The atom cloud generated by the 2D MOT forms a needle-shaped distribution aligned along the longitudinal axis. This two-dimensional cooling process produces slow atomic beams with velocities typically below $30\ \mathrm{m/s}$~\cite{dieckmann1998a}. The push beam serves to accelerate these pre-cooled atoms through the DP tubes into the science cell, following a trajectory analogous to that of a projectile. For atoms to successfully traverse both stages of the DP tubes, they must possess a sufficiently high initial longitudinal velocity to complete the passage, yet low enough to remain within the capture range of the 3D MOT. 

The push beams for both species are collimated on an optical breadboard near the Rb 2D MOT glass cell, combined using a dichroic mirror, and directed through the DP channels into the vacuum from the Rb 2D MOT side. The beam intensity, collimation, and alignment are adjusted experimentally to optimize the loading rate into the 3D MOT while minimizing any distortion to the 3D MOT.

\subsection{Dual-species 3D MOT}
The dual-species 3D MOT optics follow the same retro-reflection principle. The laser beams, emitted from the PM fiber with a numerical aperture (NA) of $0.12$, are collimated using a $f=75\ \mathrm{mm}$ lens to a waist of about $7\ \mathrm{mm}$. A dichroic mirror combines 780 nm and 671 nm laser beams. In the horizontal plane, the combined beam is then divided into two sub-beams, which cross at the center of the glass cell at an angle of 90 degrees. The vertical beam follows the same combination scheme and is further directed upwards using a right-angled elliptical mirror for vertical confinement. The required $\sigma^+/\sigma^-$-circular polarizations relative to the magnetic field direction were generated using broadband polarizing beam splitters (PBS) and dichroic quarter-wave plates. The 3D MOT coil, which consists of an anti-Helmholtz coil pair, generates the quadrupole field. This coil delivers the magnetic field gradient of $\partial B_z/\partial z=15.2 \ \mathrm{G/(cm\cdot A)}$ in its axial direction.

\section{Results and Discussions}\label{sec:results_and_discussions}
\subsection{Evaluating the Li short-distance Zeeman slower}
The fluorescence emitted by the trapped Li atoms was collected using a photodiode (PD, Thorlabs FDS10X10). A lens with $f = 50$~mm and 1-inch clear aperture was positioned at approximately $2f$ from the MOT center, collecting a small fraction of the isotropic fluorescence. The photodiode current was converted into a voltage signal via a transimpedance amplifier (TIA) with a feedback resistor $R_f = 1$~M$\Omega$. Losses along the optical path, including the glass cell, PBS, and collection lens, were calibrated to determine the effective fraction of collected light. Using this setup, the total number of trapped atoms was inferred from the measured photocurrent and the calculated photon scattering rate. With the slope of the loading curve within the initial loading process serving as a reference for the loading rate, we optimized the loading rate and the number of trapped atoms inside the 3D MOT. The dynamics of the Li 3D MOT loading with and without Zeeman slowing are demonstrated in Fig.~\ref{fig:figure4}. In dilute MOTs, one-body collisions with hot background particles predominantly determine and constrain the trap's performance. Conversely, in high-density atomic MOTs, two-body collisional processes rapidly deplete the initial high atomic flux, stabilizing the system at a steady-state equilibrium\cite{steane1992, monroe1990}. The rate equation can describe this process as:
\begin{equation}
    \frac{\mathrm{d} N}{\mathrm{~d} t}=R-\Gamma N-\beta \frac{N^2}{V},
\end{equation}
where $R$ is the loading rate, $\Gamma=1/\tau$  is the loss rate coefficient arising from background collisions, and $\beta$ is the two-body collisional loss coefficient. Solving this differential equation with the initial condition, $N_0=0$ yields the expression of the MOT dynamics (assuming isotropic density distribution)\cite{dinneen1999}
\begin{equation}
    N(t) = N_{ss} \left( \frac{1 - e^{-\gamma t}}{1 + \xi e^{-\gamma t}} \right),
    \label{eq:mot_loading}
\end{equation}
with $\beta'=\beta/V$, $\gamma = \Gamma + 2 \beta' N_{ss}$ and $\xi = \beta' N_{ss}/(\Gamma + \beta' N_{ss})$. The loading rate can be inferred as
\begin{equation}
    R=\Gamma N_{s s}+\beta'N_{ss}^2.
\end{equation}

\begin{figure}[t]
\centering
\includegraphics{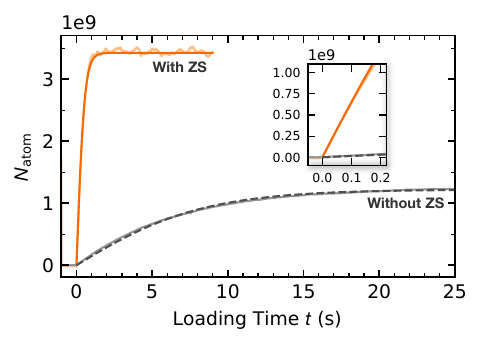}
\caption{\label{fig:figure5}
Fluorescence measurement of the Li 3D MOT loading process with and without Zeeman-slower enhanced atomic flux.
Data were obtained at an oven temperature of $T_\mathrm{Li}=372\,^\circ\mathrm{C}$.
Orange solid curve: with Zeeman slower, with a loading rate of $R^\mathrm{Li}_\mathrm{zs}=6.639(25)\times10^9\,\mathrm{atoms/s}$.
Gray dashed curve: loading without Zeeman slower, with a loading rate of $R^\mathrm{Li}_\mathrm{no~zs}=1.506(23)\times10^8\,\mathrm{atoms/s}$.
The errors given here represent numerical fitting uncertainties.
The gain factor of the Li Zeeman slower is $G = R^\mathrm{Li}_\mathrm{zs}/R^\mathrm{Li}_\mathrm{no~zs} \approx 44$.
The loading curves are fitted with the MOT loading model in Eq.~\ref{eq:mot_loading}.
}
\end{figure}

The Zeeman slower significantly boosts the atomic flux, thereby enhancing both the loading rate and the saturation atom number. Specifically, the loading rate increases by a factor of 44 to $R^\mathrm{Li}_\mathrm{zs} = 6.639(25) \times 10^{9}~\mathrm{atoms/s}$ at a moderate oven temperature of $T_\mathrm{Li} = 372~^\circ\mathrm{C}$.  This temperature is $80~^\circ\mathrm{C}$ lower (23 times lower vapor pressure) than typical $^{6}\mathrm{Li}$ experiments using traditional Zeeman slowing techniques~\cite{wu2017LiK}, which also contributes to prolonging the lifetime of the $^{6}\mathrm{Li}$ source.

\subsection{Li flux dependence}
\begin{figure}[t]
\centering
\includegraphics{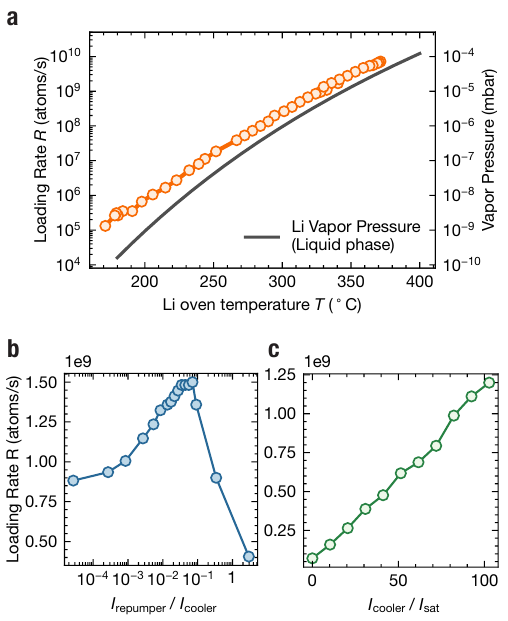}
\caption{\label{fig:figure6}
Optimization of the Zeeman slowing parameters.
(a) 3D MOT loading rate as a function of Li oven temperature.
The solid line represents the theoretical Li vapor pressure in the liquid phase~\cite{gehm2003}.
(b) Loading rate dependence on the repumper-to-cooler intensity ratio.
(c) Loading rate dependence on the Zeeman beam power.
Measurements in panels (b) and (c) were conducted at a fixed oven temperature of $335\,^\circ\mathrm{C}$.
}
\end{figure}

We systematically investigated key parameters affecting the Li flux, as shown in Figure~\ref{fig:figure6}.
First, we examined the dependence of the loading rate on the Li oven temperature (Fig.~\ref{fig:figure6}a). As expected, the loading rate increases correspondingly with oven temperature due to the elevated background vapor pressure and subsequently the atom density. Up to $370~^\circ\mathrm{C}$, no saturation behavior was observed, with the loading rate reaching a maximum of $R^{Li} \sim 1 \times 10^{10}~\mathrm{atoms/s}$ at this temperature. However, operating at higher oven temperatures accelerates the Li deposition on the in-vacuum mirror, increases pressure in the Li octagon chamber, and degrades vacuum quality in the science cell. 

To optimize overall system performance, we selected a routine working temperature of approximately $330~^\circ\mathrm{C}$, which balances loading efficiency with the preservation of the Li source lifetime, mirror optical quality, and vacuum conditions required for subsequent experimental stages. This moderate temperature setting also provides sufficient time for concurrent Rb loading. Nevertheless, our results demonstrate the significant potential of the combined 2D MOT and short-distance Zeeman slower configuration for applications demanding higher loading rates and faster experimental cycles.

Next, we investigated the influence of the ZS repumping beam's frequency and power. As described in subsection~\ref{subsection:li_laser_system}, the repumping beam is generated via an EOM operating at approximately $228~\mathrm{MHz}$. Frequency tuning of the EOM across the range of $178-245~\mathrm{MHz}$ revealed no substantial impact on the loading rate, although the optimal sideband-to-carrier ratio exhibited minor shifts throughout this range. In contrast, the power ratio between the ZS cooling and repumping beams demonstrated a pronounced influence on system performance, as illustrated in Fig.~\ref{fig:figure6}b. The loading rate initially increased with modulation depth of the EOM, reaching a maximum at $I_\mathrm{repumper}/I_\mathrm{cooler}\approx 0.07$. Beyond this optimal point, the loading rate declined precipitously due to excessive power transfer from the carrier to the sidebands, significantly reducing the effective ZS cooling power.

During operational characterization, we observed a strong correlation between the ZS cooling beam power and the resulting 3D MOT loading rate. As depicted in Figure~\ref{fig:figure6}c, the loading rate increases linearly with ZS beam power throughout the entire range accessible with our laser system. Within these power constraints, we did not observe saturation of the loading rate, suggesting potential further enhancements with higher ZS beam powers in future implementations.
\subsection{Dependence on cooling and repumping beam intensities}
\begin{figure}[t]
\centering
\includegraphics{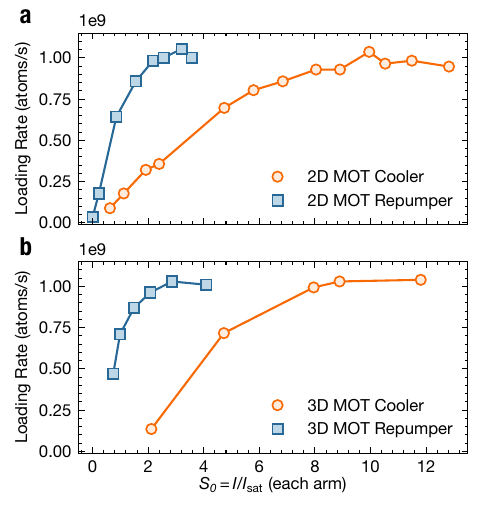}
\caption{\label{fig:figure7}
Dependence of the Li 3D MOT loading rate on cooling and repumping beam intensities.
(a) Loading rate of the 3D MOT as a function of the 2D-MOT cooling and repumping beam intensities.
(b) Loading rate of the 3D MOT as a function of the 3D-MOT cooling and repumping beam intensities.
During each measurement, all other laser intensities were kept fixed at their optimum values listed in Table~\ref{tab:tab1}.
The Li oven temperature was maintained at $335\,^\circ\mathrm{C}$ throughout the measurements.
}
\end{figure}

To systematically investigate the influence of laser power on MOT performance, we varied the cooling and repumping beam intensities for both the Li 2D and 3D MOTs while maintaining an optimized Zeeman slower configuration. Figure~\ref{fig:figure7} presents the loading rate as a function of laser intensities.

For the 2D MOT, the loading rate saturates when the cooling beam intensity reaches approximately $7 \times S_0$, while the repumping beam exhibits saturation at $2.5 \times S_0$. This yields an optimal intensity ratio of $S_{0,\mathrm{cool}}: S_{0,\mathrm{repump}} \approx 3:1$, consistent with our discussion in section~\ref{subsection:li_laser_system} regarding the necessity of strong repumping to prevent atoms from escaping the cooling cycle due to the unresolved hyperfine structure of the $|2\ ^2\mathrm{P}_{3/2}\rangle$ state in lithium. 

In the 3D MOT, loading rate saturation occurs at cooling beam intensities of $8.5 \times S_0$ and repumping intensities of $2.5 \times S_0$, resulting in an optimal ratio of $S_{0,\mathrm{cool}}: S_{0,\mathrm{repump}} \approx 3.4:1$. This ratio is comparable to values reported in similar lithium MOT setups.

\subsection{Dual-species MOT operation}
\begin{figure*}[htbp]
    \centering
    \includegraphics{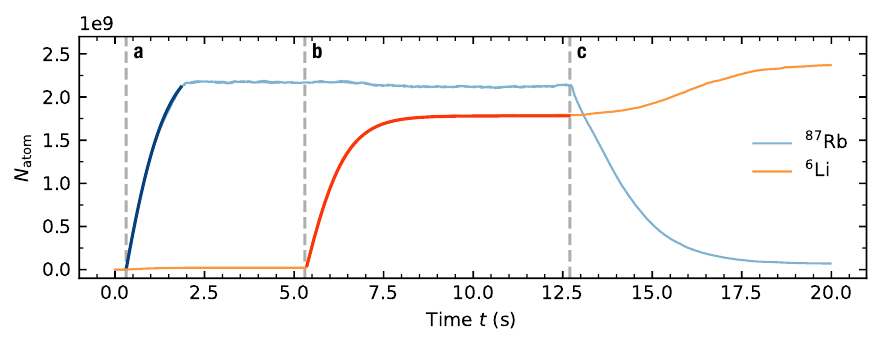}
    \caption{Dual-species MOT operation. Region a: Loading Rb by activating the Rb push beam and magnetic field (loading rate $R_\mathrm{Rb}=2.32(11)\times10^9\ \mathrm{atoms/s}$); Region b: After Rb MOT saturation, the Li push beam is activated, initiating Li loading (loading rate $R_\mathrm{Li}=1.637(16)\times10^9\ \mathrm{atoms/s}$ at an oven temperature of $T_\mathrm{oven}=350\ ^\circ\mathrm{C}$); Region c: When the Rb push beam is deactivated, the Rb MOT population decays while the Li MOT is maintained. Notably, the Li atom number increases due to reduced Rb-Li interspecies collisions. The loading curves are fitted according to Eq.~\ref{eq:mot_loading}. \textit{Note: The Rb atom number appears to saturate at $2\times10^9\ \mathrm{atoms}$ after 2 s due to the saturation of the trans-impedance amplifier output. The actual atom number exceeds this measurement limit.}}
    \label{fig:figure8}
\end{figure*}
We optimized dual-species loading by maximizing the loading rates as well as the total number of trapped atoms in steady state. The optimized laser parameters for dual-species loading are summarized in Table~\ref{tab:tab1}. Through systematic optimization, we determined that the optimal magnetic field gradient for simultaneous trapping of both species is approximately $\partial B_z/\partial z=33.4 \ \mathrm{G/cm}$, representing a compromise between the field strengths individually optimal for Li and Rb. For the detection of the dual-species MOT, we employed a dichroic mirror to separate the fluorescence signals from the atomic cloud, allowing independent analysis of each species. Fig.~\ref{fig:figure8} illustrates a loading sequence that demonstrates the successful coexistence of trapped $^6\mathrm{Li}$ and $^{87}\mathrm{Rb}$ atoms within the same trapping region. Notably, the saturated atom number for the dual-species MOT is reduced compared to single-species loading, indicating the presence of interspecies collisional losses (see Figure~\ref{fig:figure8}). The achieved loading rates and saturation atom numbers, both on the order of $1\times10^9\ \mathrm{atoms}$, provide an excellent starting point for subsequent cooling stages. This robust initial atom number is sufficient for efficiently producing double-degenerate atomic mixtures and ultimately synthesizing $^6\mathrm{Li}^{87}\mathrm{Rb}$ ground-state molecules.

\begin{table}[h]
\caption{\label{tab:tab1}
Optimized laser parameters for dual-species $^6$Li–$^{87}$Rb MOT operation.
The beam waist refers to the $1/e^2$ waist, and the detunings $\delta$ are
defined with respect to the corresponding transitions shown in
Fig.~\ref{fig:figure2}.
The EOM in the Zeeman-slower branch is driven with 15\,dBm RF power,
generating sidebands at $\pm228.28$\,MHz relative to the ZS cooling beam,
with $I_{\mathrm{repumper}}/I_{\mathrm{cooler}}\approx0.056$.
The optimal 3D MOT coil gradient is
$\partial B_z/\partial z = 33.4\,\mathrm{G/cm}$.
}
\begin{ruledtabular}
\begin{tabular}{llccc}
Species & Beam & Power (mW) & Detuning & Waist (mm) \\
\colrule
Li & 2D cooling    & 204  & $-5.0\,\Gamma_{\mathrm{Li}}$ & 14.2 \\
   & 2D repumping  & 60.5 & $-2.0\,\Gamma_{\mathrm{Li}}$ & 14.2 \\
   & ZS cooling    & 435  & $-88.1\,\Gamma_{\mathrm{Li}}$ & 9.5 \\
   & pushing       & 3.1  & $-1.2\,\Gamma_{\mathrm{Li}}$ & 0.6 \\
   & 3D cooling    & 82.5 & $-5.0\,\Gamma_{\mathrm{Li}}$ & 7.1 \\
   & 3D repumping  & 26.4 & $-2.0\,\Gamma_{\mathrm{Li}}$ & 7.1 \\
Rb & 2D cooling    & 220  & $-3.3\,\Gamma_{\mathrm{Rb}}$ & $9.4\times40$ \\
   & 2D repumping  & 8.2  & $-2.6\,\Gamma_{\mathrm{Rb}}$ & $9.4\times40$ \\
   & pushing       & 0.35 & $-3.8\,\Gamma_{\mathrm{Rb}}$ & 0.5 \\
   & 3D cooling    & 157  & $-3.3\,\Gamma_{\mathrm{Rb}}$ & 7.0 \\
   & 3D repumping  & 9.3  & $-2.6\,\Gamma_{\mathrm{Rb}}$ & 7.0 \\
\end{tabular}
\end{ruledtabular}
\end{table}

\section{Summary and Conclusion}\label{sec:summary}
In this work, we have demonstrated the implementation and optimization of a compact dual-species 2D MOT system for $^{6}$Li and $^{87}$Rb, incorporating a short-distance Zeeman slower for lithium. The setup delivers high atomic fluxes and efficient loading into a dual-species 3D MOT, providing an excellent starting point for quantum gas experiments. The short-distance Zeeman slower plays a key role in boosting Li capture efficiency while retaining lower technical complexity compared to conventional setups. Most optimization parameters for Li and Rb remain largely decoupled, enabling straightforward tuning of each species.

Our dual-species system showcases the feasibility of simultaneously trapping and cooling heavy and light atomic species with substantially different physical properties. The design principle is directly applicable to other atomic species combinations, such as lithium-cesium and sodium-potassium. 
This work represents a solid step toward compact and versatile dual-species and molecule setups for studying few-body physics and strongly correlated many-body systems with long-range dipole-dipole interactions.


During the preparation of this manuscript, we became aware of a compact high-flux ytterbium-rubidium Zeeman slower~\cite{chen2025}.

\begin{acknowledgments}
We gratefully thank P. Preiss, S. Jochim, M. Hetzel, C. Klempt, J. Zeiher, P.-J. Wang, and C. Gross for stimulating discussions, A. Mayer, and M. Antic for their technical contributions, Hao Lin Yu for contributions at the early stage of the Li laser system.
The project is funded by the European Union (ERC, DiMoBecTe, 101125173). We gratefully acknowledge support from the Max Planck Society, and the Deutsche Forschungsgemeinschaft under Germany's Excellence Strategy -- EXC-2111 -- 390814868.  

\end{acknowledgments}

\bibliography{main}
\clearpage
\appendix

\section{Differential pumping channels design}\label{sec:appendix_DP}

\begin{figure*}[t]
\centering
\includegraphics{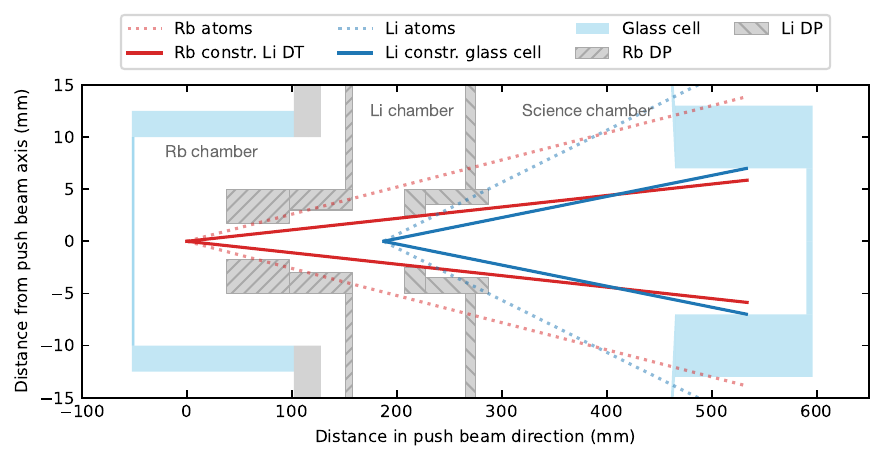}
\caption{\label{fig:s_figure4}
Design of the differential pumping stages. 
Rubidium atoms are cooled in the 2D MOT at $x=0\,\mathrm{mm}$ and pushed toward the science cell with a divergence of 26\,mrad (dotted red line, example from Ref.~\cite{holtkemeier2011}).
Lithium atoms leave the 2D MOT at $x=187\,\mathrm{mm}$ with a divergence of 50\,mrad (dotted blue line, example from Ref.~\cite{tiecke2009}) and are not further clipped by the Li differential pumping stage compared to the science cell opening angle (solid blue line). 
The maximum capturable divergence of rubidium and lithium atoms is therefore set by the opening angle of the Li DP and the science cell, respectively (solid lines).
}
\end{figure*}

It is essential to carefully design the dimensions of the differential pumping channels (DP) between the chambers to maintain a sufficient pressure ratio while not blocking the atomic flux more than necessary. The atomic beam divergence is given as its transverse-to-longitudinal velocity ratio, $\zeta = v_\mathrm{tr}/v_x$. The first is a result of the finite transverse temperature after the 2D MOT stage, while the latter can be tuned via the push beam power. In general, it is advantageous to reduce the distances from 2D MOTs to the science cell as much as possible, resulting in a larger opening angle of the science cell and thus an increased capturable fraction of atoms. The DP dimensions are then chosen to not clip the atomic beam significantly more than the science glass cell, which is shown schematically in Figure~\ref{fig:s_figure4}. The DPs consist of two sections in series (with diameters $d_1$ and $d_2$ as well as lengths $l_1$ and $l_2$), as it is easier to manufacture compared to one conical tube (as used, e.g., in~\cite{hammel2025, tobiashammel2021}). Additionally, the Rb DP is 45$^{\circ}$ polished on the surface facing the Rb atoms, oriented on previously reported designs~\cite{jollenbeck2011, holtkemeier2011, parry2015}. The resulting conductances are listed in table~\ref{tab:s_tabs4} alongside with respective DP dimensions.

\begin{table}[b]
\caption{\label{tab:s_tabs4}
Dimensions and conductances of the differential pumping (DP) channels.
Diameters $d_1$ and $d_2$ and lengths $l_1$ and $l_2$ are given in mm.
Conductances $C$ for nitrogen are given in units of l/s.
}
\begin{ruledtabular}
\begin{tabular}{lccccc}
Channel & $d_1$ & $l_1$ & $d_2$ & $l_2$ & $C$ \\
\colrule
Rb DP & 3.5 & 60 & 6 & 60 & 0.064 \\
Li DP & 5   & 20 & 7 & 60 & 0.270 \\
\end{tabular}
\end{ruledtabular}
\end{table}

\section{Zeeman Slower Beam Simulations}\label{sec:appendix_ZS}

The effectiveness of the Zeeman slower beam is influenced by both the laser configurations (intensity and detuning) and the magnetic field configuration for the 2D MOT (the number of magnets and the spacing between magnet stacks). While these parameters can, in principle, be optimized experimentally, practical constraints, such as chamber manufacturing, make this process challenging. The laser detuning is limited by the tuning range of the AOM, while the spacing of the magnets is constrained by the fixed dimensions of the octagon chamber. Moreover, adjusting the number of magnets necessitates removing and reinserting entire magnet holders, which in turn requires realignment after each modification. Given these constraints, it is highly advantageous to perform simulations beforehand to identify optimal starting conditions for the experiment.

\subsection{Numerical principle}\label{appendix_ZS_principle_numerical}

The core of the simulations for the Zeeman slower beam is the numerical integration of the equations of motion obtained from the position and velocity-dependent radiation force acting on the atoms (cf. Eq.~\ref{eqn:zeeman_slower_radiation_force}), using the python package \texttt{scipy.intergrate version 1.15.0}. The atoms' starting position is the bottom of the oven, \SI{190}{mm} away from the 2D MOT center. The magnetic field is calculated with \texttt{magpylib version 4.3.0}, with the measured magnetization of \SI{8.8e5}{A/m} from~\cite{tiecke2009}. 

As the gain (cf. Eq.~\ref{eqn:zeeman_slower_gain}) strongly depends on the 2D MOT capture velocity itself, the latter has been simulated along the $z$-direction, in an analogous way as for ZS, with two effective counter-propagating MOT beams in the region $|z|\leq \sqrt{2}  \times w_{0,\mathrm{2DMOT}}$:
\begin{align}
    \delta_{\pm} &= \delta \mp vk \pm \frac{\mu' \eta_B B}{\hbar}, \label{eqn:lirb_mot_detuning} \\
    \label{eqn:mot_radiation_force}
    F_{\pm} &= \pm \frac{\hbar k \Gamma}{2} \left( \frac{S_0}{1 + S_0 + \left(2 \delta_{\pm}/\Gamma \right)^2} \right), \\
    a_{\text{tot}} &= \frac{1}{m} \left( F_{+} + F_{-} \right) \times \sin(\alpha) \ ,
\end{align}
where $\alpha = 42.5^\circ$ is the angle between the horizontal plane and the MOT beams. To obtain the full trajectories in Fig.~\ref{fig:figure4}, the forces from equations~\ref{eqn:zeeman_slower_radiation_force} and~\ref{eqn:mot_radiation_force} are added, and the resulting equations of motion are again integrated following the same procedure. \\

Different $x-y-$displacements from the main axis affect the gain via three primary aspects: (1) variations in the magnetic field profile along the z-axis, (2) the $x-z$-dependent Gaussian intensity of the Zeeman beam (ZB), and (3) changes in the capture velocity arising from both the magnetic field profile and the Gaussian beam's shape and displacement. Given the small solid angle of the 2D-MOT region relative to the oven, purely vertical trajectories from the oven to the MOT are considered, with equal probability across different displacements.

Throughout the calculation, a perfect two-level system (TLS) is assumed, without any losses. Accounting for losses goes beyond the scope of this simplified model, which aims to provide a relative estimation of the expected performance under different parameter settings.

\begin{figure*}[t]
\centering
\includegraphics{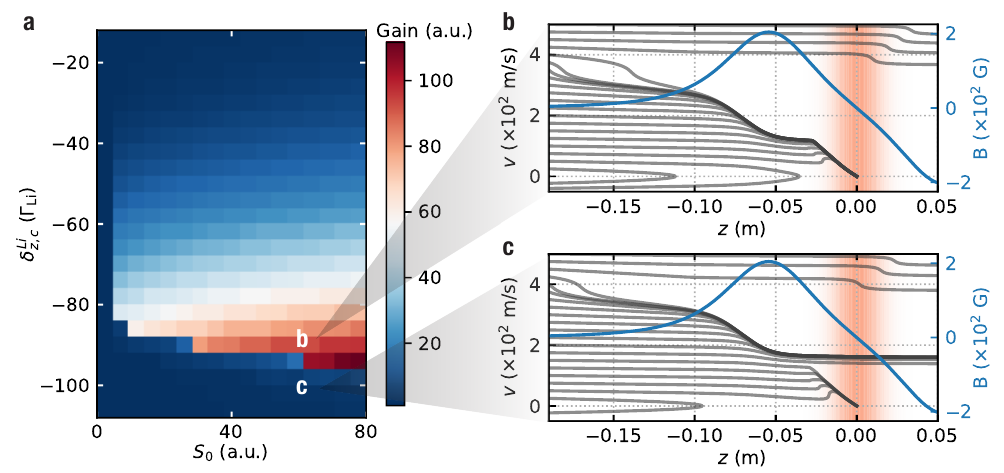}
\caption{\label{fig:s_figure1}
Zeeman-beam gain as a function of detuning and power.
The parameters are $\eta_B=0.8$, $l_x=70\,\mathrm{mm}$, $l_z=107\,\mathrm{mm}$,
$n_\mathrm{MPS}=12$, and $d_x=d_y=0$.
(a) Gain map of the Zeeman slower.
White labels in (a) indicate the laser parameters used for the trajectory
plots in (b) and (c).
(b) and (c) Atomic trajectories along the applied magnetic field.
The Gaussian beam regions of the 2D MOT are shown in orange.
In (b), the final velocities of a large fraction of atoms are reduced below
the capture velocity $v_c$ and can be trapped.
In (c), the final velocities of most atoms remain above the capture range of the 2D MOT.
}
\end{figure*}

\subsection{Parameter scan for optimization}\label{appendix_ZS_principle_optimize}

To narrow down the number of variables in the parameter scan and thereby reduce computational effort, certain parameters are investigated and fixed in advance.\\

A 2D scan of the gain over the saturation parameter $S_0$ and detuning $\delta_{z,c}^{Li}$ (Fig.~\ref{fig:s_figure1}a) verifies that higher $S_0$ leads to a higher possible gain, justifying working with a conservative guess of maximum available power. There is always a certain suitable detuning range for each intensity, where the maximum flux can be reached for detunings where $v_\mathrm{end}$  is only slightly smaller than the capture velocity $v_c$ of the MOT, as shown in figure~\ref{fig:s_figure1}b. As $v_c$ is reduced for off-axis trajectories, we conservatively continue the calculations with the detuning fixed to $v_\mathrm{end} = 1/2\,  v_c(x_0=y_0=0)$ for each parameter set. \\

The magnet stack distance in the $x$-direction, $l_x$, needs to be optimized within a range of 50 to 75 mm, constrained by the dimensions of the vacuum setup. To maintain a magnetic field gradient of up to $50\ \mathrm{G/cm}$ in the 2D MOT center \footnote{The maximum gradient is limited by the maximum number of magnets that can fit in one stack (12).}, the number of magnets per stack $n_\mathrm{MPS}$ is adapted for the different $x$-distances, as detailed in table~\ref{tab:s_tabs1}. The $z$-distance, $l_z$, is set to $107\ \mathrm{mm}$ (close to the minimum possible due to the chamber dimensions) with range $\pm 4\ \mathrm{mm}$ to allow for fine-tuning of the MOT center. As the shielding factor of the metal chamber is yet unclear to us, we scan $\eta_B$ from 0.7 to 0.9. Finally, for each of these parameter sets, the scan is carried out for displacements $dx$, $dy$ up to $9~\mathrm{mm}$ from the main axis. A summary of all scan parameters is provided in table~\ref{tab:s_tabs1}. 

\begin{table}[h]
\caption{\label{tab:s_tabs1}
Magnetic-field gradient at the center of the 2D MOT for different
magnet separations $l_x$ and numbers of magnets per stack $n_\mathrm{MPS}$.
}
\begin{ruledtabular}
\begin{tabular}{cccc}
$l_x$ (mm) & $n_\mathrm{MPS}$ & $\nabla B$ (G/cm) \\
\colrule
50 & 9  & 51.1 \\
55 & 10 & 50.0 \\
60 & 12 & 50.8 \\
65 & 12 & 45.9 \\
70 & 12 & 41.4 \\
75 & 12 & 37.2 \\
\end{tabular}
\end{ruledtabular}
\end{table}

\begin{table}[h]
\caption{\label{table:tab:s:tabs2}
Scan parameters for Zeeman-slower optimization.
For each scan, the other laser parameters are fixed as listed in
Table~\ref{tab:tab1}, and the magnet parameters are fixed as listed in
Table~\ref{tab:s_tabs1}.
}
\begin{ruledtabular}
\begin{tabular}{lcc}
Description & Parameter & Range \\
\colrule
Laser detuning & $\delta_0$ & cf.\ Fig.~\ref{fig:s_figure2}(a) \\
Magnet stacks $x$ distance & $l_x$ & 55--75\,mm \\
Magnet stacks $z$ distance & $l_z$ & 103--111\,mm \\
Magnetic-field shielding factor & $\eta_B$ & 0.7--0.9 \\
Main-axis displacement & $d_x, d_y$ & 0--9\,mm \\
\end{tabular}
\end{ruledtabular}
\end{table}

\begin{figure}[h]
\centering
\includegraphics{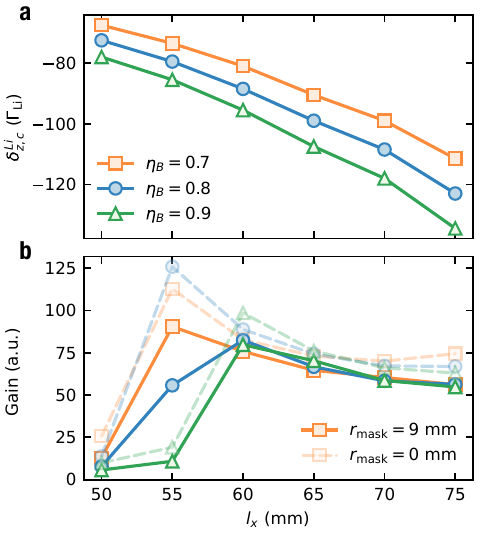}
\caption{\label{fig:s_figure2}
Optimized detuning (a) and gain (b) as a function of the magnet $x$-distance $l_x$.
The detunings in (a) were obtained by setting the final velocity $v_f = 1/2\,v_c$.
The radius $r_\mathrm{mask}$ of a circular mask is centered on the MOT, and the average gain is computed over this area.
Variations of $l_z$ and $r_\mathrm{mask}$ in the range 3--6\,mm led to changes smaller than 10\% and are not plotted for clarity.
}
\end{figure}

The simulation results indicate that a magnet stack distance of $l_x \ge \SI{60}{mm}$ is advisable, as on-axis gains over 60 are observed in this regime,  regardless of the different magnetic field shielding factors. Neither the magnet displacement in z-direction of \SI{4}{mm}, nor the choice of radius around $(dx, dy) = (0,0)$ changes the gain significantly. 

\subsection{Extended simulation results for final experimental parameters}

Based on the parameter scan, the magnet distance was fixed at $l_x = 70\ \mathrm{mm}$, providing a good compromise between gain and convenience during setup installation. Starting with the suggested value, the detuning was then fine-tuned in the experiment to maximize the gain within the AOM tuning range. With the optimized laser parameters from table~\ref{tab:tab1}, the final simulated gain averaged over an area is listed in table~\ref{tab:s_tabs3}. For the average over a circular area with a radius of $r_\mathrm{mask}=9~\mathrm{mm}$ and $\eta_B = 0.8 $, the gain yields 53. 

\begin{figure*}[h]
\centering
\includegraphics{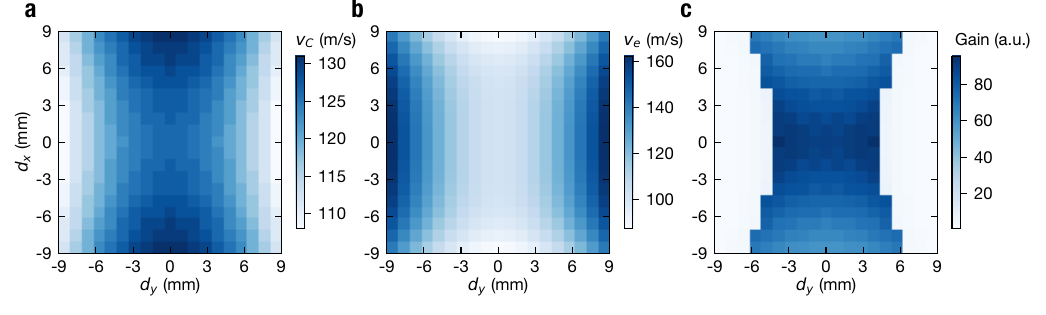}
\caption{\label{fig:s_figure3}
(a) Capture velocity of the bare 2D MOT, $v_c$; 
(b) main final velocity after Zeeman slowing, $v_e$; and 
(c) gain as a function of $x$-$y$ displacement. 
The main final velocity is defined as the point where $\partial v_f(v_0) / \partial v_0$ is minimal (see Figure~\ref{fig:figure4}a, inset). 
The gain is greater than one only in regions where $v_e < v_c$. 
The remaining simulation parameters for the Zeeman beam are: $\eta_B=0.8$, $l_x = 70$ mm, $l_z = 107$ mm, $n_\mathrm{MPS} = 12$, $\delta^{Li}_{z,c} = -86.7\,\Gamma$, and $S_0 = 60.4$.
}
\end{figure*}

\begin{table}[h]
\caption{\label{tab:s_tabs3}
Gain values for different magnetic-field shielding factors $\eta_B$
and averaging radii $r_\mathrm{mask}$.
The remaining Zeeman-beam parameters are
$\eta_B=0.8$, $l_x=70\,\mathrm{mm}$, $l_z=107\,\mathrm{mm}$,
$n_\mathrm{MPS}=12$, $\delta^{Li}_{z,c}=-86.7\,\Gamma$,
and $S_0=60.4$.
}
\begin{ruledtabular}
\begin{tabular}{lcccc}
$\eta_B$ &
\multicolumn{4}{c}{$r_\mathrm{mask}$ (mm)} \\
\cline{2-5}
 & 0 & 3 & 6 & 9 \\
\colrule
0.7 & 5.3 & 4.8 & 3.5 & 3.4 \\
0.8 & 89  & 88  & 74  & 53  \\
0.9 & 69  & 68  & 68  & 61  \\
\end{tabular}
\end{ruledtabular}
\end{table}
The corresponding capture velocities and the gain as a function of $x-y$ displacement are shown in Figure~\ref{fig:s_figure3}. However, it is essential to note that simplifications are made in our model, particularly the assumption of a perfect TLS and the use of a 1D model, which affects both the Zeeman beam and the capture velocity simulation. The accuracy of the simulated value to the experiment, thus, may result from several uncertainties (actual capture velocity due to imperfections, the accurate value of $\eta_B$, losses to different hyperfine states, atom trajectories not strictly along $z$, etc.) acting on top of each other. 

\end{document}